\documentclass[12pt,showpacs,showkeys,preprintnumbers,nofootinbib]{revtex4-1}

\usepackage{url}
\usepackage{ulem}

\usepackage{geometry}
\usepackage{amssymb,amsmath,graphics,graphicx}  

\usepackage{rotate,color,colortbl}
\usepackage[dvipsnames]{xcolor}
\usepackage[latin1]{inputenc}

\usepackage{subcaption}

\usepackage{tikz}
\usetikzlibrary{shapes,arrows}

\begin{document}

\newcommand{\pderiv}[2]{\frac{\partial #1}{\partial #2}}
\newcommand{\deriv}[2]{\frac{d #1}{d #2}}

\title{ Optimal diffusion in ecological dynamics \\ with Allee effect in a metapopulation }

\author{Marcelo A. Pires}
\thanks{piresma@cbpf.br}
\affiliation{Brazilian Center for Research in Physics, Brazil}

\author{S\'ilvio M. Duarte~Queir\'os}
\thanks{sdqueiro@cbpf.br}
\affiliation{Brazilian Center for Research in Physics, Brazil}
\affiliation{
National Institute of Science and Technology for Complex Systems}

\date{\today}

\begin{abstract}
How diffusion impacts on ecological  dynamics under the Allee effect and spatial constraints? That is the question we address.
Employing a  microscopic minimal model in a metapopulation (without imposing nonlinear birth and death rates) we evince
 --- both numerically and analitically --- the emergence of an optimal diffusion that maximises the  survival probability. Even though, at first such result seems counter-intuitive, it has empirical support from recent experiments with engineered bacteria. Moreover, we show that this optimal diffusion disappears for loose spatial constraints.
\end{abstract}


\maketitle

\tableofcontents

\section{Introduction}

The Allee effect, the influential finding named after the ecologist Warder Clyde Allee \cite{Allee1991}, is phenomenon typically manifested by a departure from the standard logistic growth that enhances the susceptibility to extinction of an already vulnerable sparse population. Curiously, W.~C. Allee did not provide a definition of the effect~\cite{Stephens_Sutherland_Freckleton1999}, but in general terms it can be defined as \textit{"the positive association between absolute average individual fitness and population size over some finite interval."}~\cite{JMDrake_AMKramer2011}. 
The strong Allee effect, which is the focus of this work, corresponds to the case when the deviation from the  logistic growth includes an initial population threshold below which the population goes extinct~\cite{FCourchamp_LBerec_JGascoigne2008}. On the other hand weak version of the Allee effect treats positive relations between the overall individual fitness and population density and does not present threshold population size nor density.

The 
Allee effect can turn up from a variety of mechanisms such as  mate limitation, cooperative breeding, cooperative feeding, habitat amelioration ~\cite{JMDrake_AMKramer2011,FCourchamp_LBerec_JGascoigne2008}. 
Although empirical support for the Allee Effect is little, it is possible to find instances thereof in some terrestrial arthropods, aquatic invertebrates, mammals, birds, fish, and reptiles~\cite{FCourchamp_LBerec_JGascoigne2008,AMKramer2009}. Despite that fact, there is nowadays enough technology that allows Synthetic Biology to program new collective behaviour in bacteria, including the Allee effect~\cite{RSmith2014}. 

Besides Ecology and Conservation Biology~\cite{FCourchamp_LBerec_JGascoigne2008}, there is a growing number of studies addressing the importance of the Allee Effect in other subjects such as Epidemiology \cite{RRRegoes_DDieter_SBonhoeffer2002,ADeredec_FCourchamp2006,FMHilker_MLanglais_HMalchow2008} and Cancer Biology \cite{KSKorolev_JBXavier_JGore2014,LSewalt2016} among others. Explicitly, in Ref.~\cite{KSKorolev_JBXavier_JGore2014} the authors suggest  the manifestation of the Allee Effect as the tumor growth threshold may be explored in therapeutics.

For long the Allee effect was mostly studied at the population scale, but in Ref.~\cite{PAmarasekare1998} it was shown the relevance of this effect at the metapopulation level as well. Afterwards, it was effectively demonstrated the Allee effect at the metapopulation level can emerge from the Allee effect at the local population level~\cite{Zhou_ShuRong_GangWang2004,Zhou_ShuRong_GangWang2006}.

 Focussing on the theoretical approach to the problem, several  models ranging from phenomenological to purely microscopic have been able to successfully capture the Allee effect and to explore its dynamical consequences~\cite{Boukal_Berec2002,Berec2008}, namely those coping with the interplay between the Allee Effect and dispersal. Let us mention some examples hereinafter: on the one hand, there are works showing a positive association between migration and the number of invaded patches \cite{Ackleh_Allen_Carter2007}; the invasion diagram presented in Ref.~\cite{Keitt_Lewis_Holt2001} shows that the propagation failure regime shrinks as the dispersal rate increases, whereas in Ref.~\cite{Brassil2001} it as asserted that an increase in migration leads to an increase in the mean time to extinction in a simple metapopulation dynamics.  
On the other hand, there are works indicating that combination of the Allee effect and dispersal negatively impacts population dynamics as in Ref.~\cite{Hopper_Roush1993} where the authors claim that the vulnerability to extinction increases with the mean-square displacement.
Considering a nonlinear dynamics analysis of the effect, the survival-extinction bifurcation diagram shown in Ref.~\cite{Hadjiavgousti_Ichtiaroglou2004} reveals 
an increase in the extinction regime as the dispersal probability increases.
The results conveyed in Ref.~\cite{Robinet2008} indicate that populations under the Allee effect face an inverse relation between the establishment probability and the  pre-mating dispersal. Complementary, it was also found that a dispersive population under the Allee effect faces a dramatically slowed spreading, especially the early spread~\cite{Veit_Lewis1996}.

Particularly in Population Ecology, Windus and Jensen~\cite{windus2007} proposed a minimal model that successfully captures the Allee Effect by means of a  bistable dynamics that naturally arises from their microscopic rules. Inspired by their model we develop an ecological metapopulation dynamics in order to  explore how the threefold interplay between the Allee Efect, diffusion and spatial constrains 
impacts on the survival probability of a population dynamics. 
It is reasonably expected that the diffusion has a beneficial impact on the population survival by decreasing the local competition for resources, but interestingly we observe that for severe spatial constrains there is the emergence of an optimal diffusion rate that promotes the highest survival probability. This nonmonotonic relation between survival and dispersal --- which is not very intuitive at first glance --- was recently observed in controlled experiment with engineered bacteria~\cite{RSmith2014}.

The remaining of this manuscript is organised as follows: in the next section  we describe our model, algorithm and the mathematical approach, respectively. In section $3$, we introduce our results and in Section $4$ we present our final remarks and future avenues of research on this subject-matter.

\section{Model and Monte Carlo Simulation}

Consider a metapopulation~\cite{Hanski1991,Hanski1998} with $L$ subpopulations composed of agents that are able to move, die or reproduce. 
As usual in metapopulation dynamics \cite{Hanski1991}, we assume a well-mixed subpopulation, i.e., inside each subpopulation all individuals have the possibility to interact
with each other.
\footnote{In Statistical Physics parlance that is to say that our local dynamics exhibits a mean-field character.}  The mobility is implemented as a random walk between the neighbour subpopulations and it occurs with probability $D$ for each agent.
At a given time step, if the diffusion event is not chosen  (probability $1-D$) then one of the two events is chosen \cite{windus2007}: death of an agent  with probability $\alpha$ or reproduction with probability $\lambda$ when two agents meet.

At this point, three remarks are worth making: first, heed that $D$ controls the time scale between migration or death/reproduction; second, it is clear that we make no extra assumptions on the probabilities  $\alpha$ or  $\lambda$; and third, this proposal naturally incorporates the environmental changeability since the carrying capacity of each subpopulation is not fixed.  Moreover, there is no local condensation  of the agents because the random walk uniforms the agents distribution among the subpopulations.

We would like to stress that our goal is not to model a specific ecological dynamics, but rather to investigate the possible emerging scenarios from this minimal agent-based  migration-reproduction-death dynamics. This approach can be naturally seasoned with further elements that account for the traits of a given system. It is well-known that the use of minimal models are very helpful in providing an understanding of the cornerstone mechanisms  present in tailored models.

\subsubsection*{Monte Carlo Algorithm}

Computationally\footnote{Our main code is availabe at  \cite{piresMA_github}.}, we use an array with $N$ states divided into the $L$ subpopulations.
 Each state in the subpopulation $u$ indicates an agent, $i_A^u$ or a vacancy, $i_V^u$. The time is measured in Monte Carlo Steps (mcs) that consists of a visit to each one of the $N$ states.

\vspace*{0.30cm}
\textbf{Monte Carlo Step:}

For each state $i=1,\ldots,N $:
\begin{itemize}
\item First get the subpopulation, say $u$, of the state $i$.
\item With probability  $D$:
 \begin{itemize}
\item \textbf{Diffusion:} If the state $i$ indicates an agent, $i_A^u$, then move  it to one of its neighbours $w$ chosen at random: $i_A^u \Rightarrow i_A^w $
  \end{itemize}

\item With probability $1-D$:
\begin{itemize}

\item \textbf{Reproduction:} If the state $i$ indicates a vacancy, $i_V^u$, then pick at random another state $j$ in the same subpopulation $u$. If this $j$ indicates an agent, $j_A^u$,
then pick at random another state $l$ in the same subpopulation $u$. If the state $l$ indicates another agent, $l_A^u$, then transform the  vacancy $i_V^u$ into an agent  $i_A^u$ with rate $\lambda$:
 $i_V^u+j_A^u+l_A^u \Rightarrow   i_A^u+j_A^u+l_A^u $

\item \textbf{Death:} If the state $i$ indicates an agent, $i_A^u$, then transform it into a vacancy with rate $\alpha$:
 $i_A^u \Rightarrow i_V^u$

  \end{itemize}
  
  \end{itemize}

After each Monte Carlo Step we apply a synchronous updating of the states.

\subsubsection*{Mathematical approach}

Consider that $A_u(t)$ and $V_u(t)$ are 
the number of agents and vacancies in the subpopulation $u$ at instant $t$, respectively. We use a ring  metapopulation where each node is a subpopulation connected to $k$ neighbour  subpopulations. The  parameter $k$ controls the magnitude of the spatial constraints. Let $u={1,\ldots,L}$. Considering the well-mixed population (mean-field) at the local scale, the time evolution of the coupled system is given by

\begin{equation}
 \frac{dV_u}{dt}
 =
 (1-D)  
 \Big[
\overbrace{-\frac{\lambda V_u A_u^2}{(V_u+A_u)^2}}^\text{Reproduction}
+
\overbrace{\alpha A_u}^\text{\newline Death}         
\Big] 
\label{Eq:Vu}             
\end{equation}

\begin{equation}
  \frac{dA_u}{dt}
 =
 (1-D)  
 \Big[
\underbrace{-\frac{\lambda V_u A_u^2}{(V_u+A_u)^2}}_\text{Reproduction}
+
\underbrace{\alpha A_u}_\text{\newline Death}         
\Big] 
+
D\Big[
\underbrace{-A_u}_\text{Emigration}
+
\underbrace{\sum_{z=1}^{L} \frac{1}{k} W_{uz} A_{z}}_\text{Immigration}
\Big]              
\label{Eq:Au}
\end{equation}

with $W_{uz}$ being the elements of the adjacency matrix which assumes the value $1$ if $u$ and $z$ are connected or 0 otherwise. The variability of the carrying capacity of each subpopulation is represented in  Eqs.~(\ref{Eq:Vu})-(\ref{Eq:Au}) by the term $V_u(t)+A_u(t)$.

Aiming at taking into account both the cases of single and multiple sources of invasion, we shall use an initial condition given by 
\begin{equation}
A_u(0)
=
 \begin{cases} 
\frac{1}{n_s} 
\frac{N}{L}  
  &  
 u={1,\ldots,n_s}
\\
0  
&   
 u={n_s,\ldots,L}
\end{cases}
\label{Eq:InitialCondition}
\end{equation}
where $N/L$ is the initial size of each subpopulation and $n_s$ is the number of initial sources. By default, we use $V_u(0)=N/L-A_u(0)$ as well.

\subsubsection*{Survival-extinction phase transition}

From a preliminary numerical analysis we observed that the steady-state solution satisfies

\begin{equation}
A_u^\infty=\bar{A}, \quad V_u^\infty=\bar{V} \quad
 \forall u \quad u={1,2,\ldots,L}
\end{equation}

Using that observation as an ansatz to solving our equations implies
\begin{equation}
N=\sum_{u=1}^{L}(A_u+V_u)=L(\bar{A}+\bar{V})
\Rightarrow
\bar{V} = N/L - \bar{A}
\end{equation}
\begin{equation}
 \frac{d\bar{A}}{dt}
 = 
 (1-D)\left[ 
         \frac{\lambda (N/L - \bar{A}) \bar{A}^2}{(N/L)^2} 
        - \alpha \bar{A} 
             \right]  +
             D\left[ - \bar{A}  + \frac{1}{k}\left(k\bar{A}  \right) \right] 
             = 0
\label{EqA2}
\end{equation}

From Eq.~(\ref{EqA2}) we can obtain three solutions to $\bar{A}$. The stability analysis yields

\begin{equation}
 A_u^\infty
=
 \begin{cases} 
\frac{N}{2L}
\left( 1\ +\sqrt{1-4\frac{\alpha}{\lambda}} \right)
  &  
A_u(t=0) \geq A_c^o \text{ and } \alpha \geq \lambda/4
\\
0  
&   
\text{otherwise}
\end{cases}
\label{EqA3}
\end{equation}

Where $A_c^o$ is the threshold initial population size  required for the local persistence:

\begin{equation}
 A_c^o  = \frac{N}{2L}
\left( 1-\sqrt{1-4\frac{\alpha}{\lambda}} \right)
\label{Eq:thresholdAo}
\end{equation}

Equations~(\ref{EqA3})-(\ref{Eq:thresholdAo}) do not explicitly take into account the diffusion parameter $D$, but they allow us to get an insight into the nature of the survival-extinction phase transition: they show that the subpopulation faces a discontinuous transition at the critical point $\alpha_c=\lambda/4$.  As in the long-term,  the mobility spreads the absence of local correlations to the whole metapopulation, then the global dynamics undergoes an abrupt phase transition as well; we numerically confirm in the next section.

\section{Results and Discussion}

\begin{figure}[!hbtp]
\centering
\includegraphics[scale=0.39]{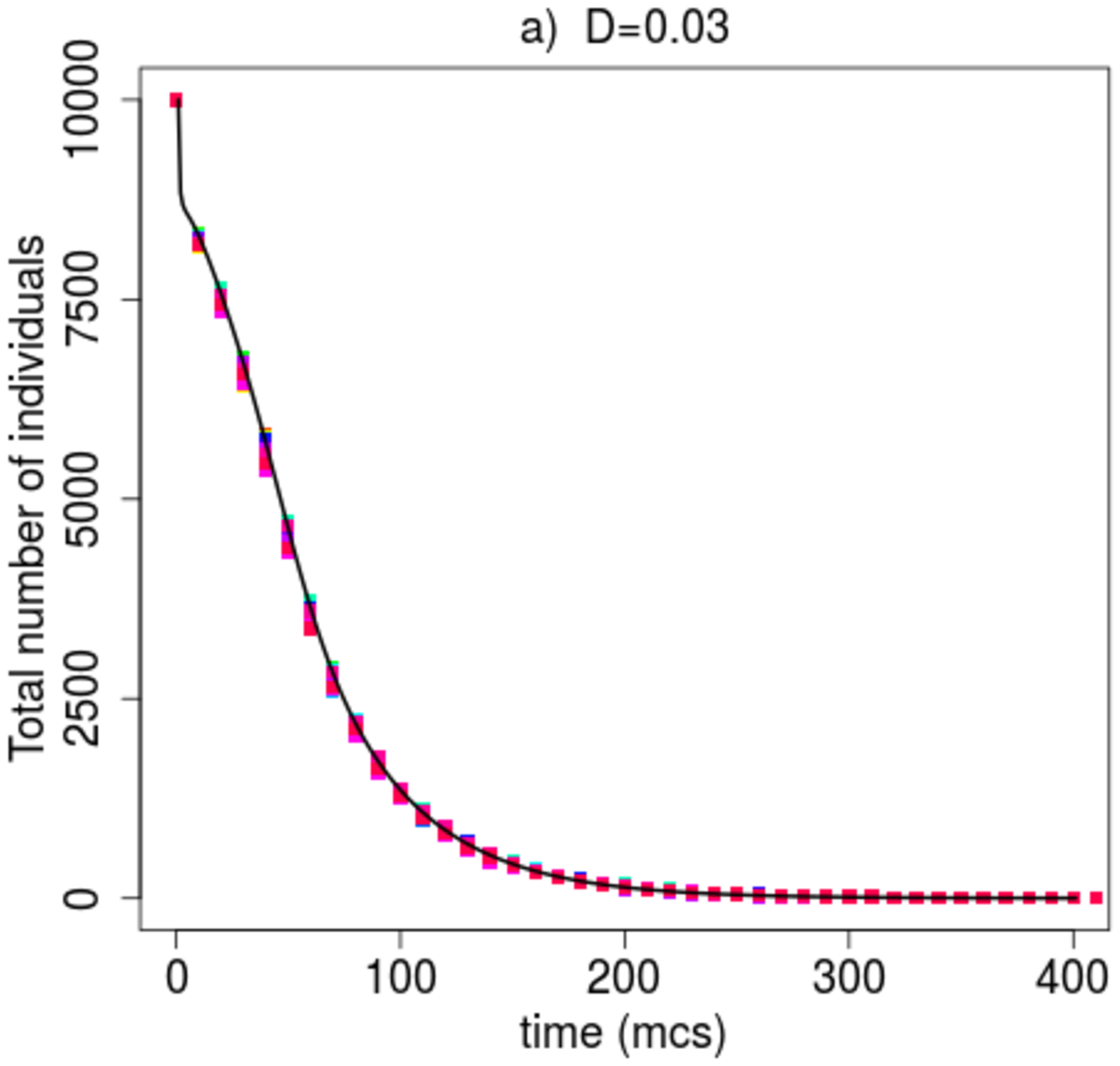}
\includegraphics[scale=0.39]{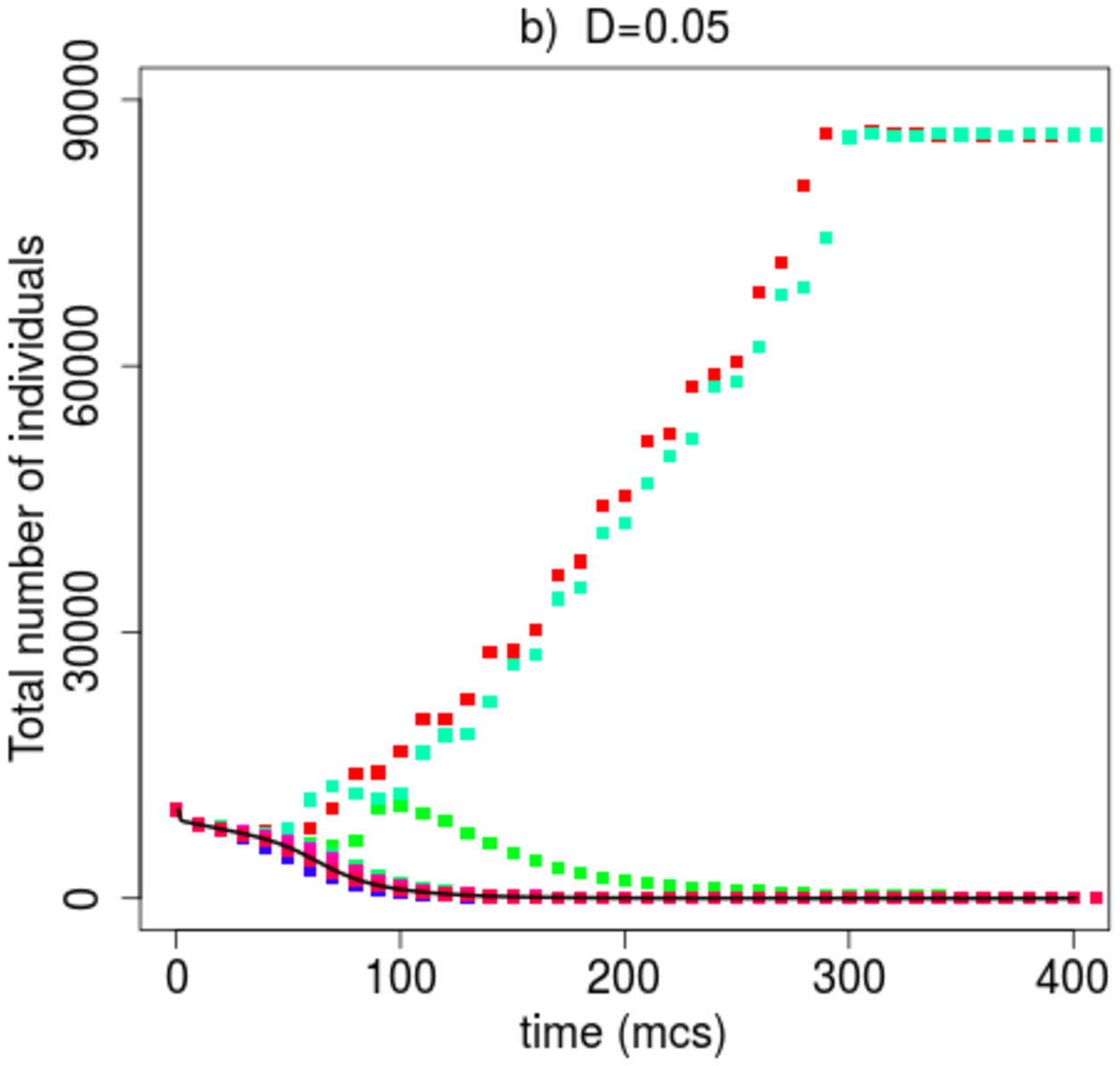}
\includegraphics[scale=0.39]{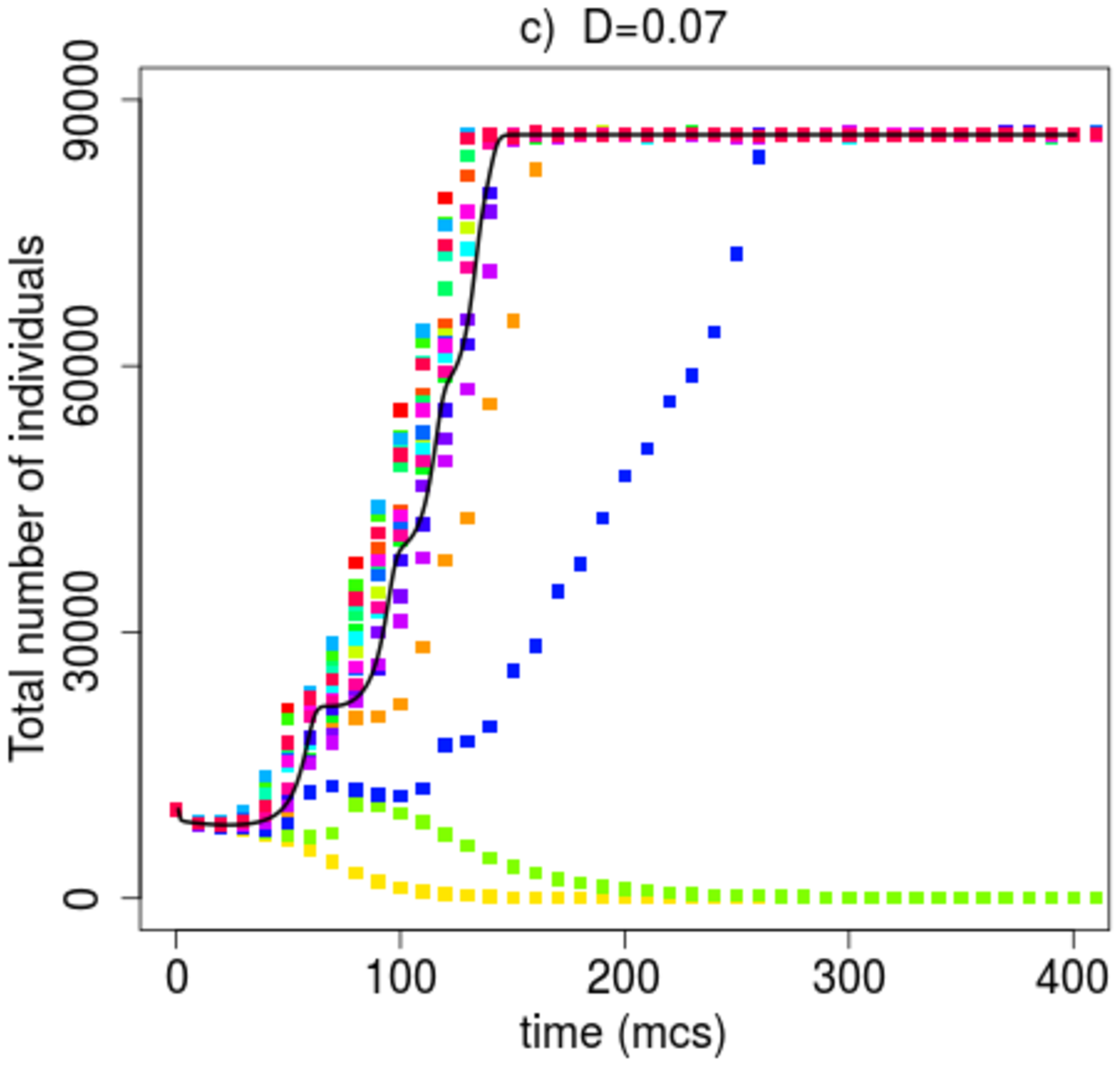}
\includegraphics[scale=0.39]{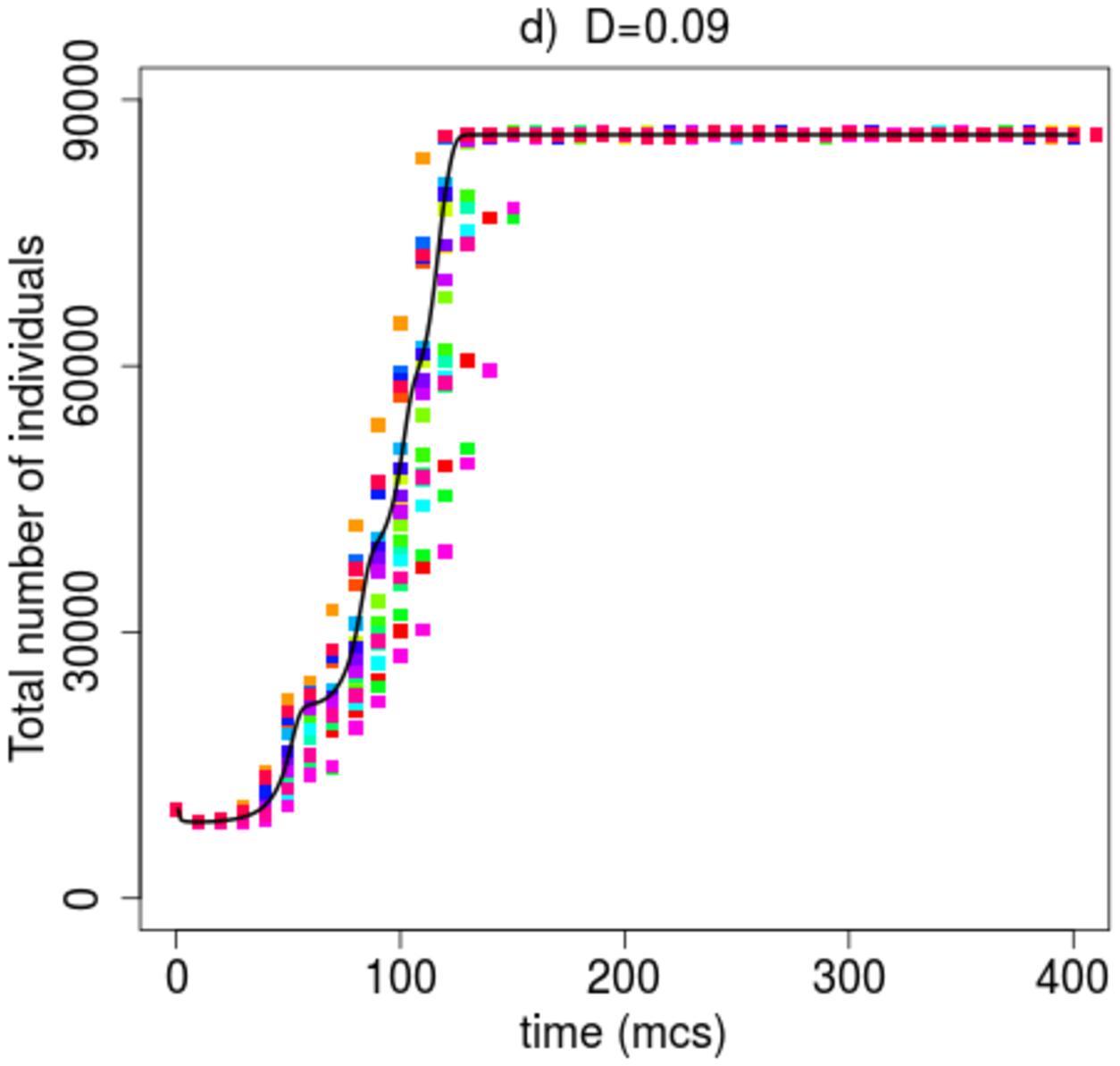}
\includegraphics[scale=0.39]{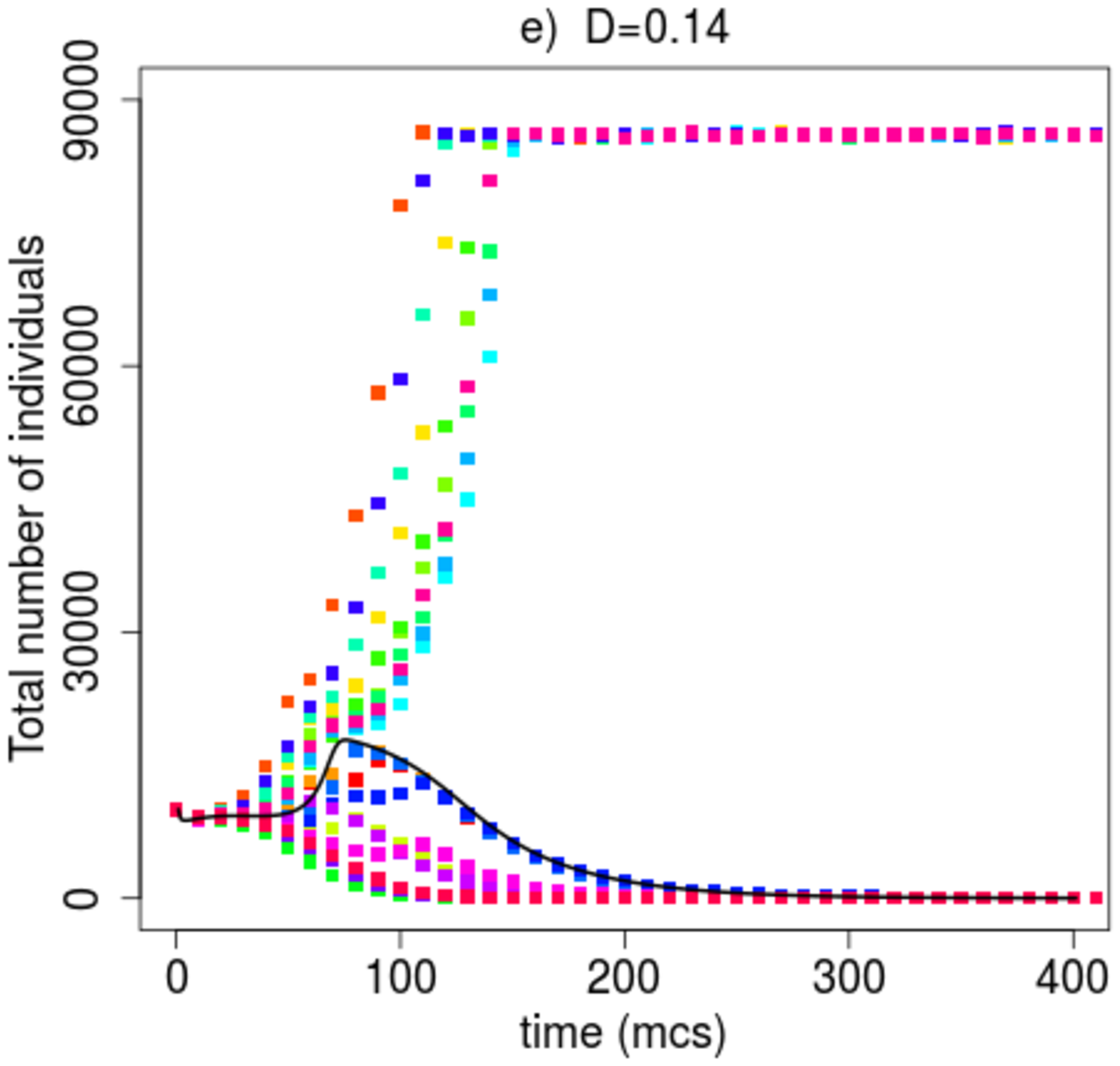}
\includegraphics[scale=0.39]{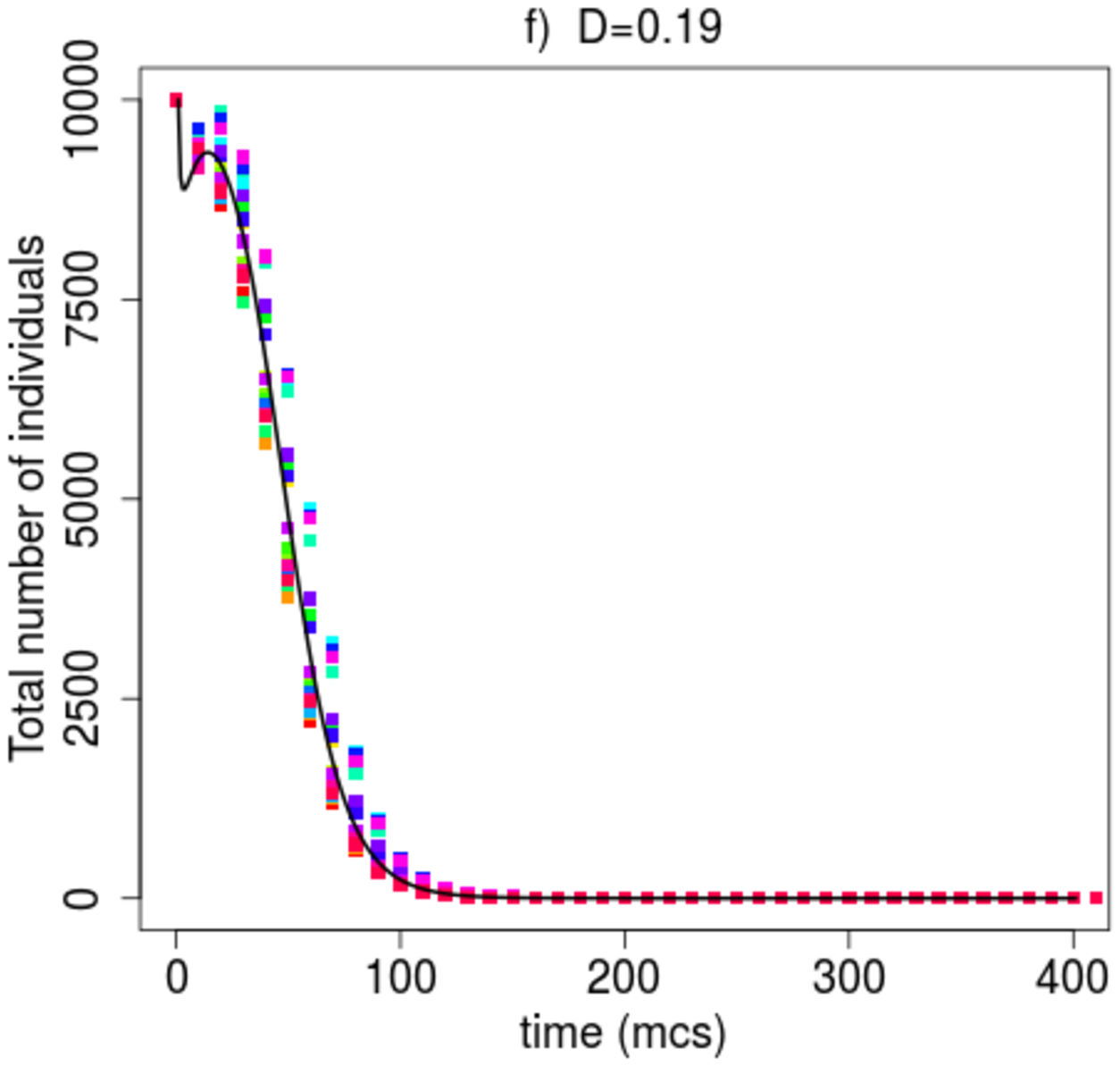}
\caption{Total number of agents vs time (in mcs) for  $D=\{0.03,0.05,0.07,0.09,0.14,0.19\}$ with $L=10$, $N=10^4 L$, $n_s=1$. The symbols were obtained from Monte Carlo simulations and the lines from Eqs.~(\ref{Eq:Vu})-(\ref{Eq:Au}).}
\label{Fig:timeseries}
\end{figure}

\begin{figure}[!hbtp]
\centering
\includegraphics[scale=0.50]{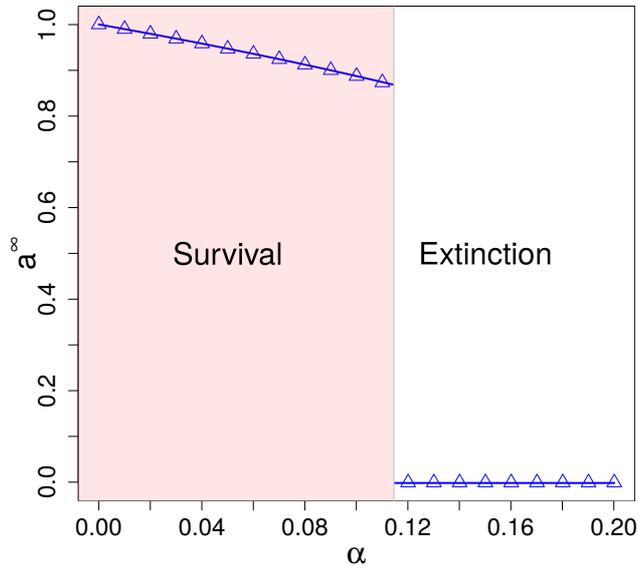}
\caption{Stationary density of agents $a^\infty$ vs mortality rate $\alpha$ with $D=0.2$, $k=2$, $L=10$, $N=10^4 L$, $n_s=1$. The symbols come from the Monte Carlo Simulations and the lines come from the  numerical integration of Eqs\ref{Eq:Vu}-\ref{Eq:Au}.}
\label{Fig:Avsalpha_1}
\end{figure}

\begin{figure}[!hbtp]
\centering
\includegraphics[scale=0.33]{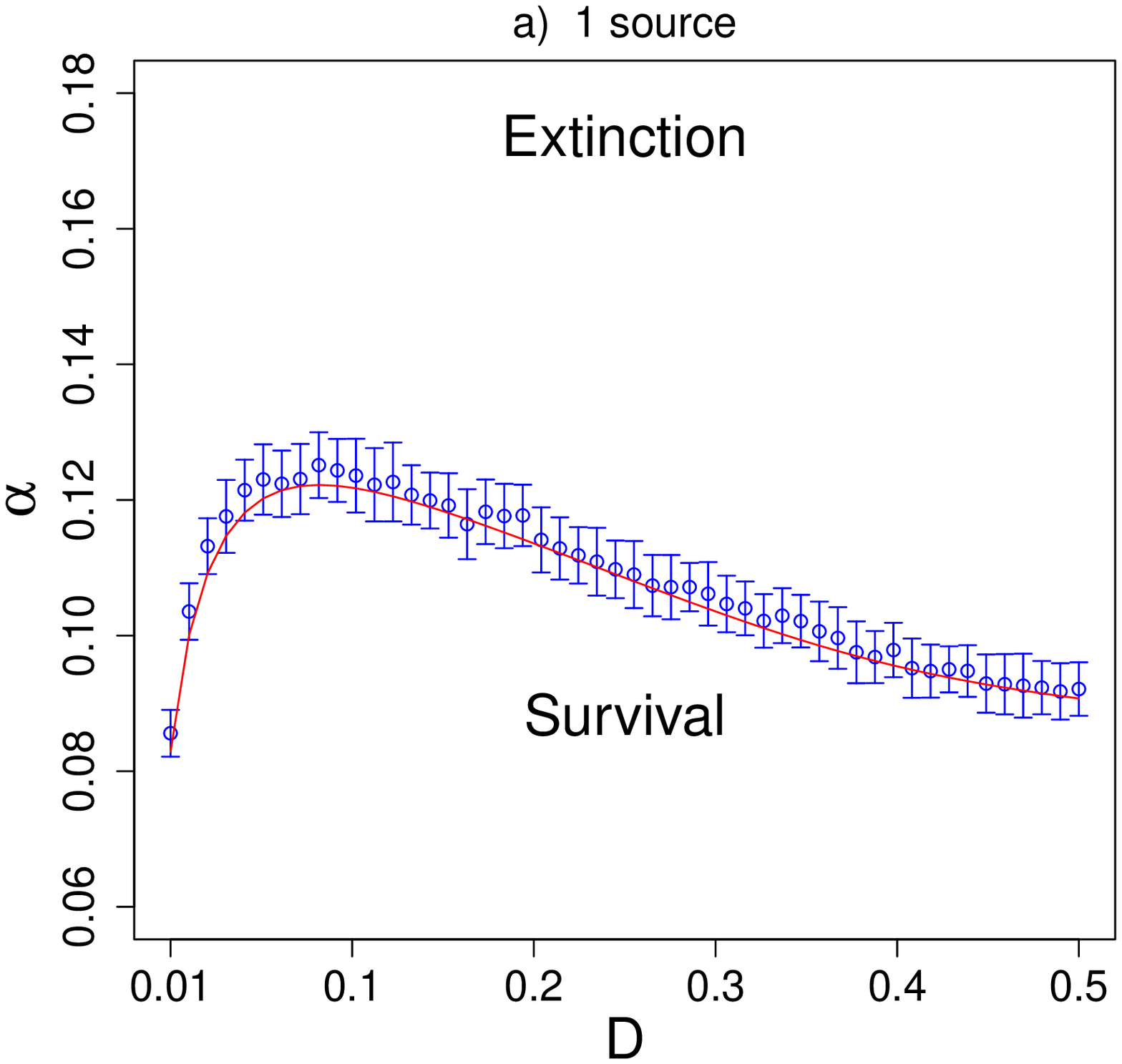}
\includegraphics[scale=0.33]{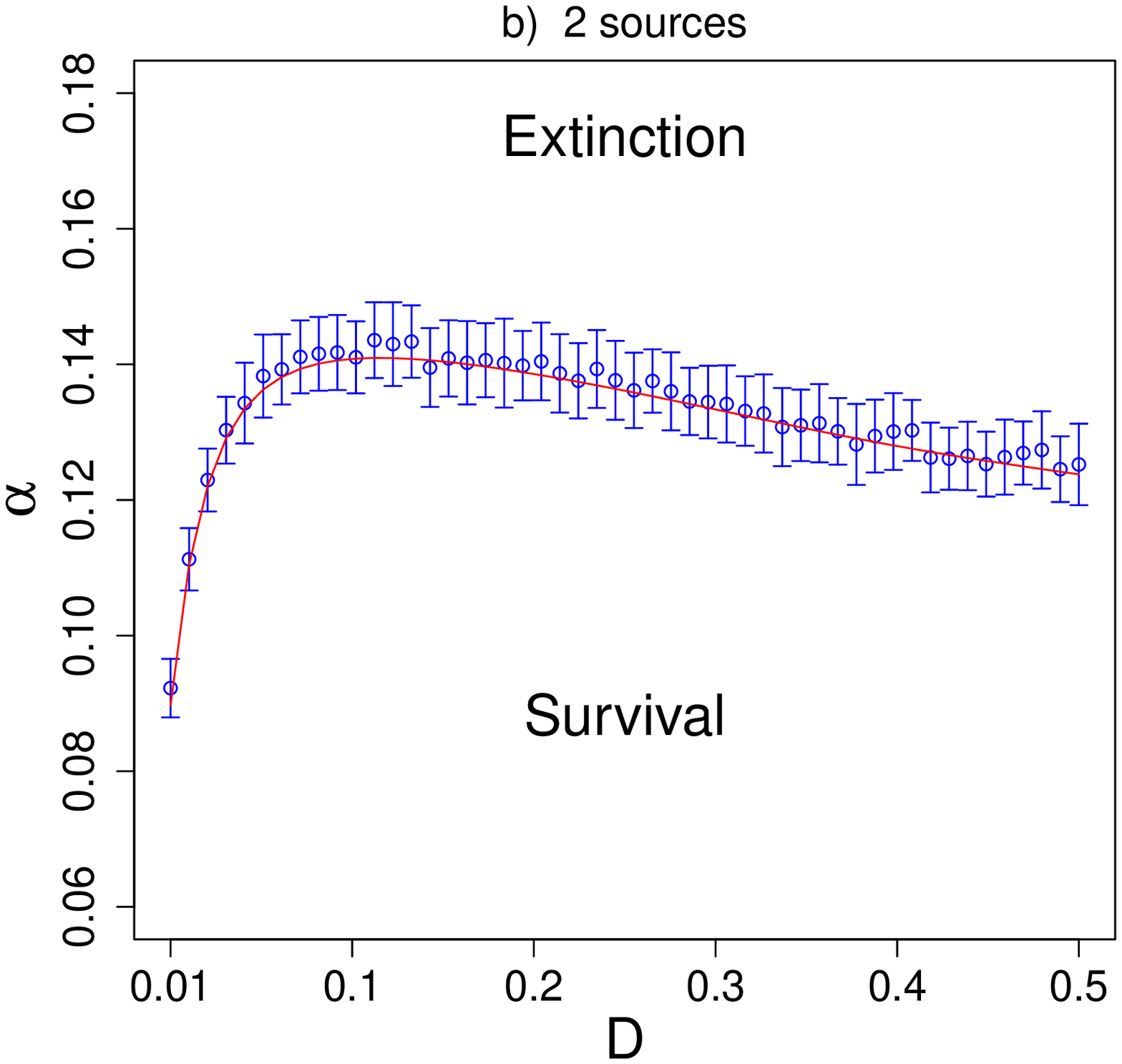}
\includegraphics[scale=0.33]{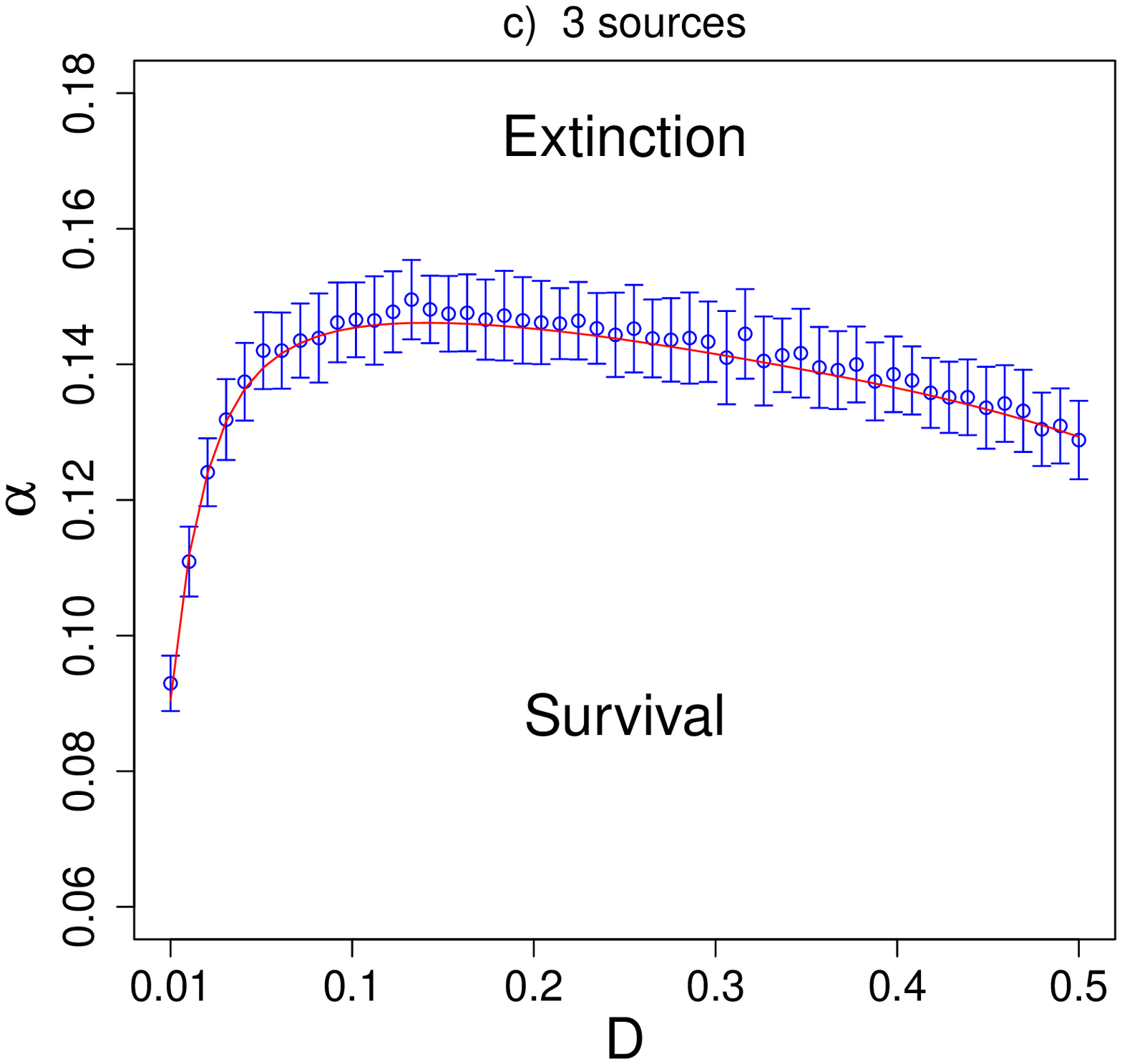}
\includegraphics[scale=0.33]{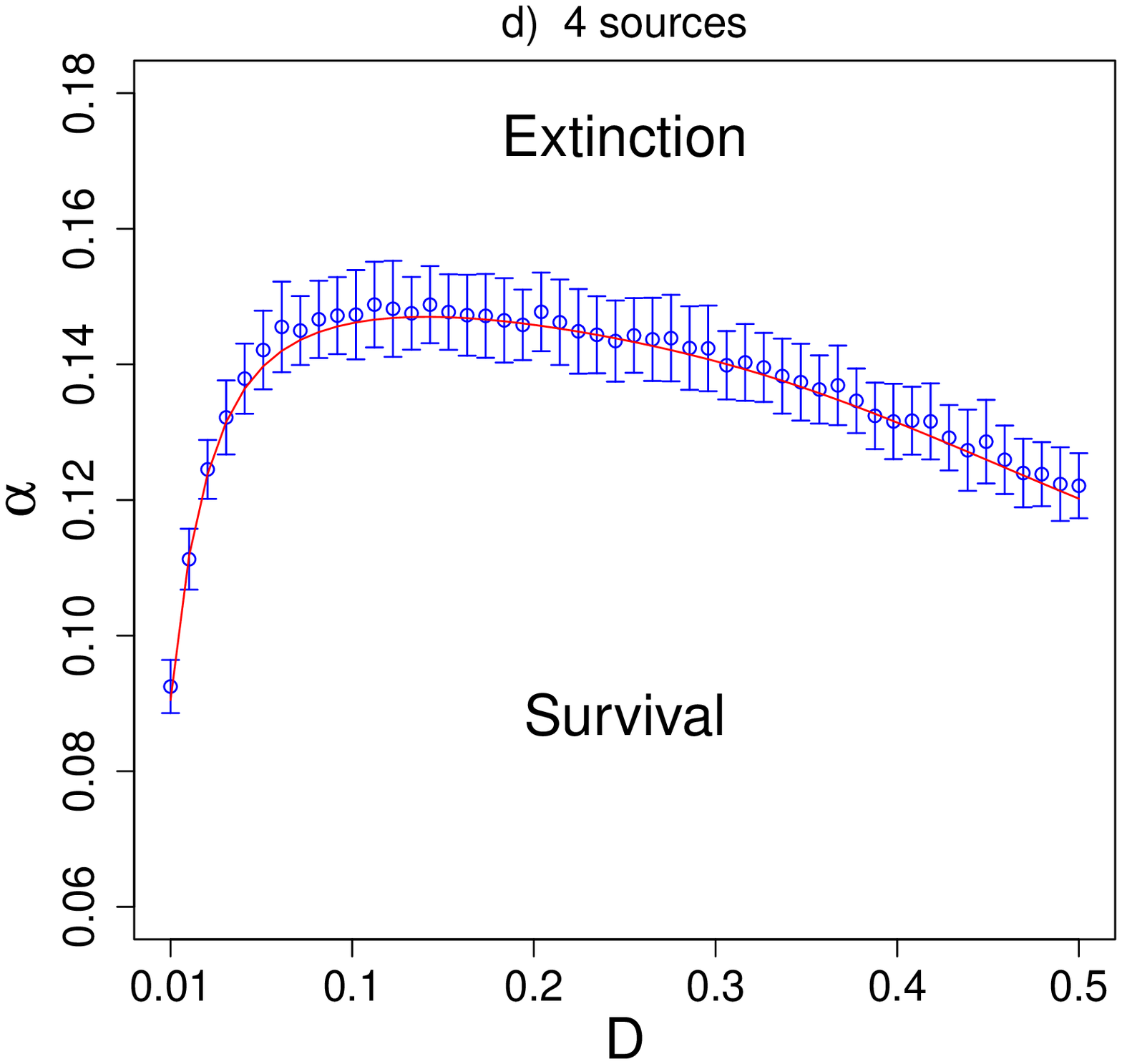}
\includegraphics[scale=0.33]{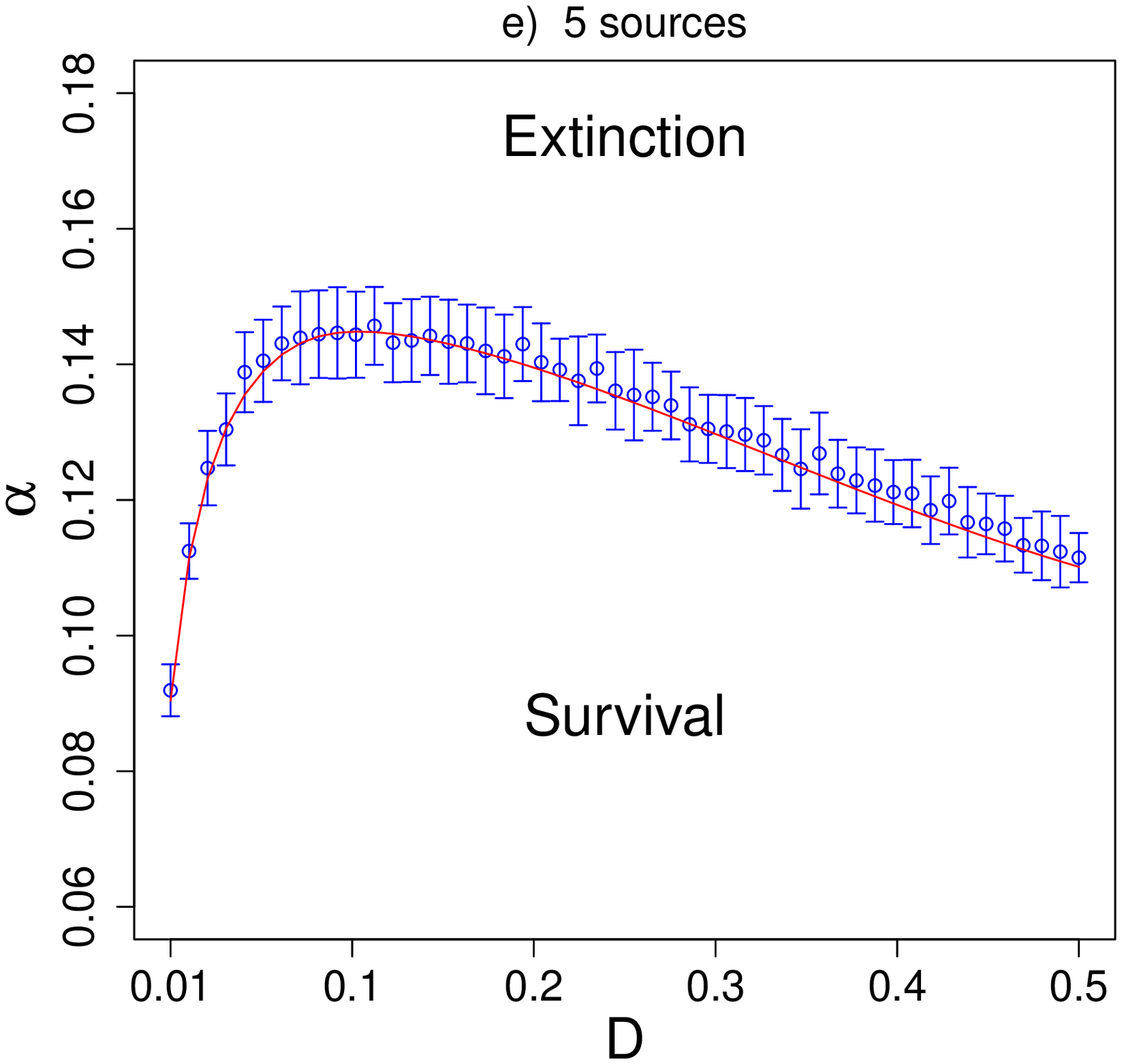}
\includegraphics[scale=0.33]{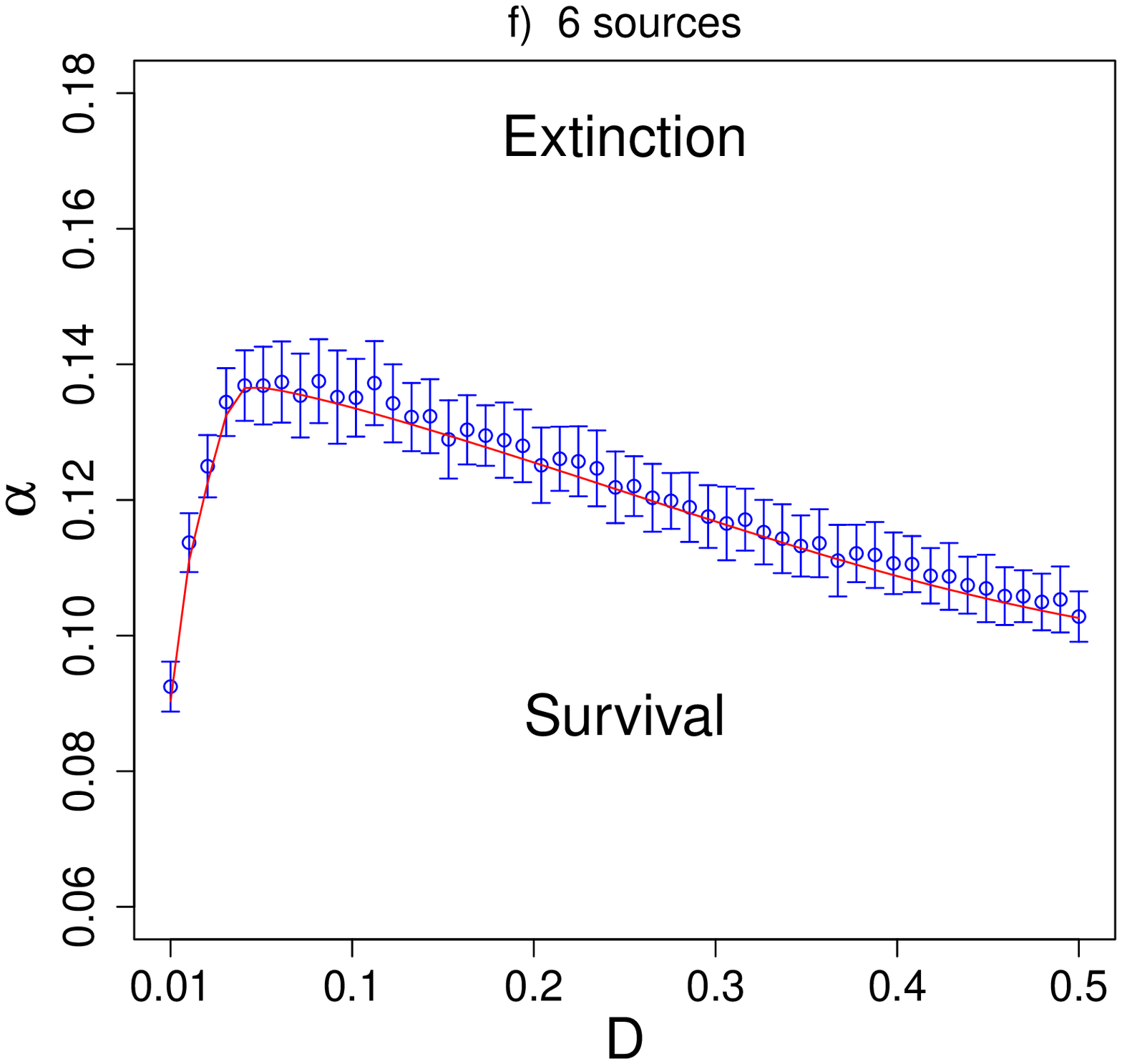}
\caption{ 
Phase diagram $\alpha $ v $D>0$ for $n_s={1,2,\ldots,6}$ sources. The point $D=0$ is excluded from the diagram since it refers to isolated populations with threshold $\alpha_c=\lambda/4=0.25$ $L=10$, $N=10^4L$. In all the cases $n_0=\frac{10^4}{n_s}$, where $n_0$ is the initial subpopulation size. The lines are obtained from Eqs.~(\ref{Eq:Vu})-(\ref{Eq:Au}).}
\label{Fig:alpha_vs_D_ns}
\end{figure}

\begin{figure}[hbtp]
\centering
\includegraphics[scale=0.5]{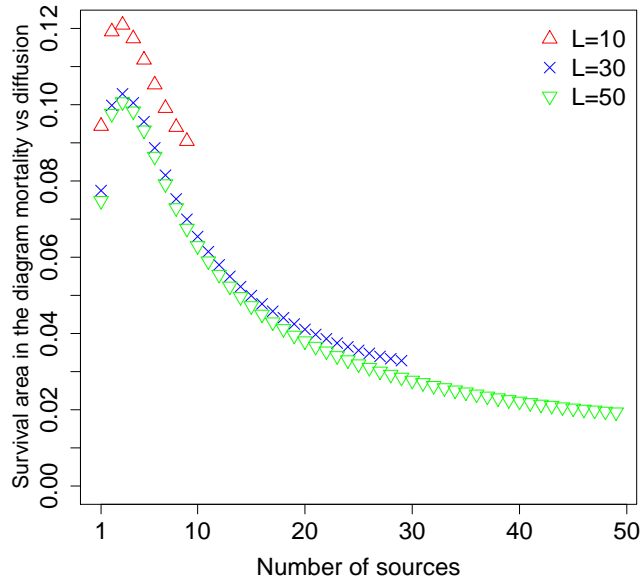}
\caption{ 
Survival area in the phase diagram $\alpha \times D$  versus the number of  sources  $n_s$ for $0<D<1$. The case $D=1$ is excluded because it implies no reproduction/death. The case $D=0$ is excluded because it implies no migration between the patches. In all cases we keep the initial subpopulation size fixed $n_0=\frac{10^4}{n_s}$ and we use $N=10^4L$. }
\label{Fig:survival_area_1}
\end{figure}

\begin{figure}[!hbtp]
\centering 
\includegraphics[scale=0.35]{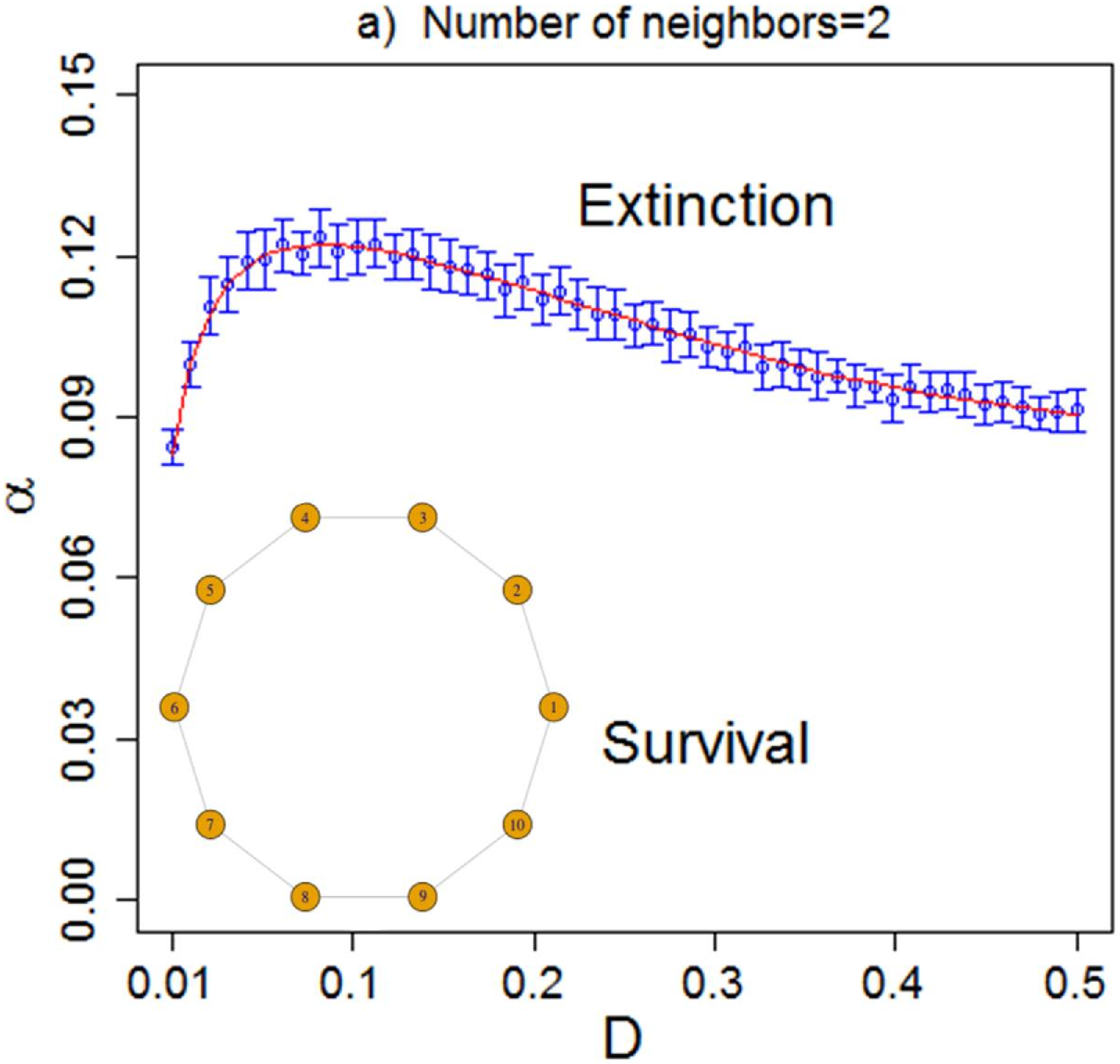} 
\includegraphics[scale=0.35]{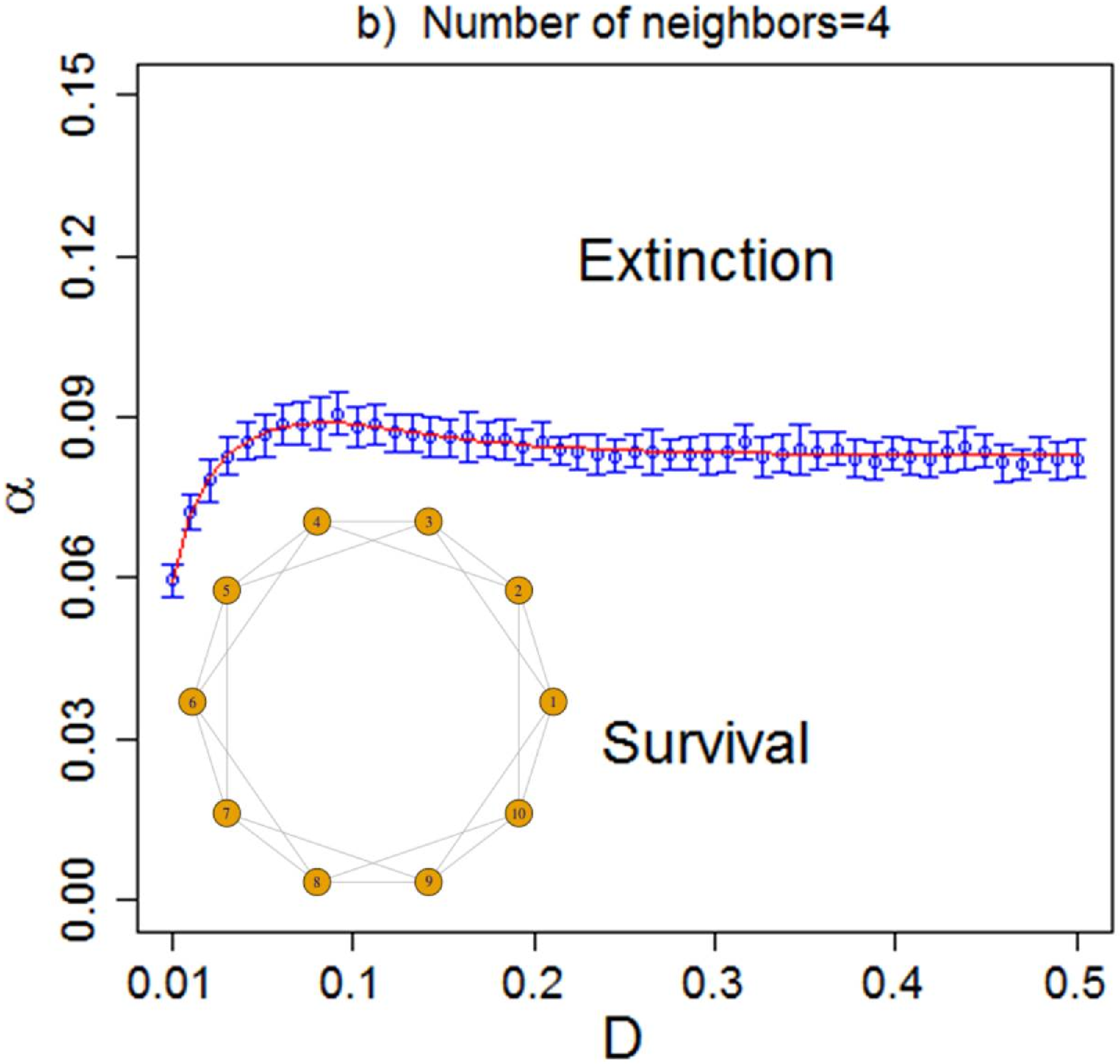}

\includegraphics[scale=0.35]{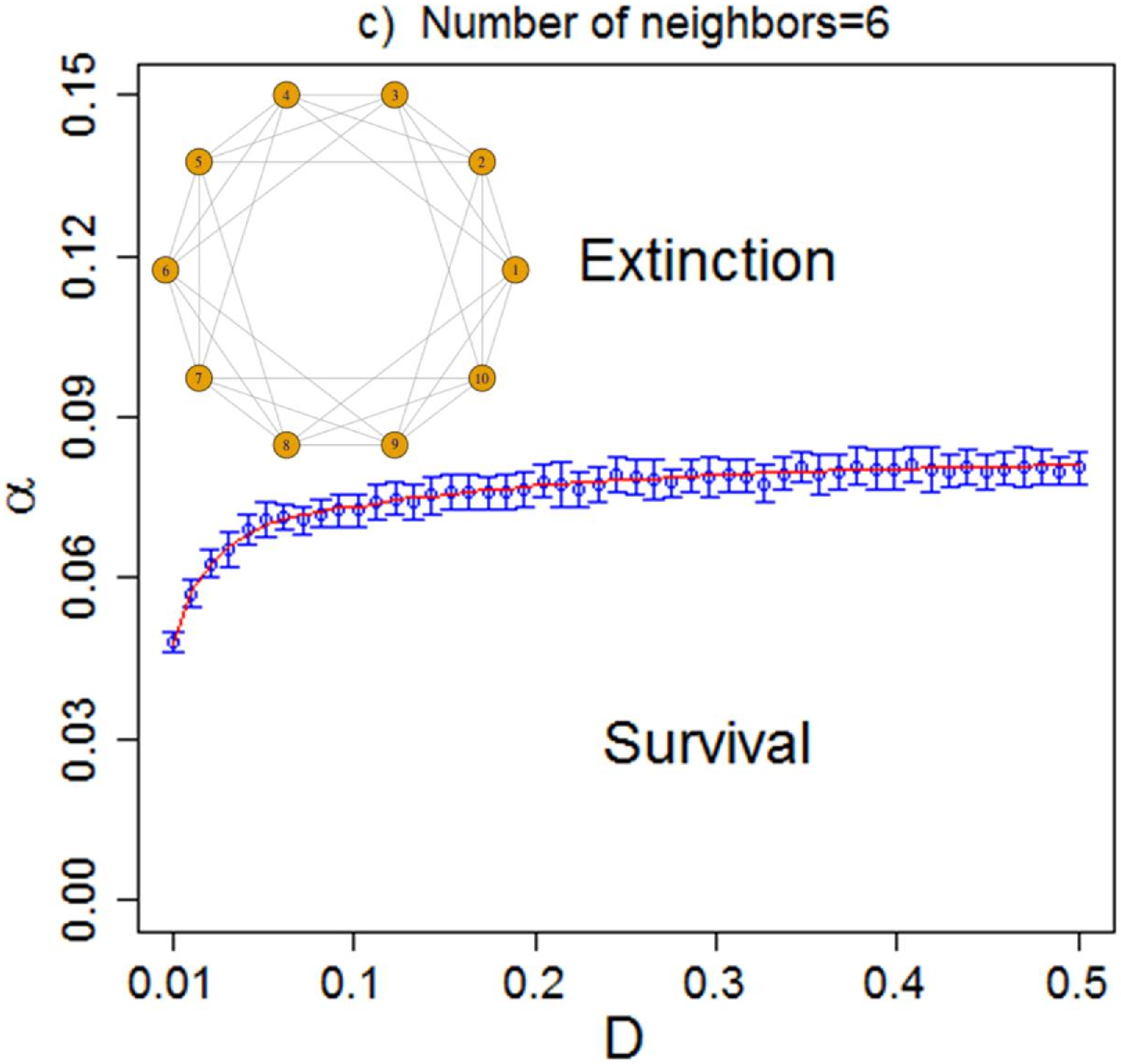}
\includegraphics[scale=0.35]{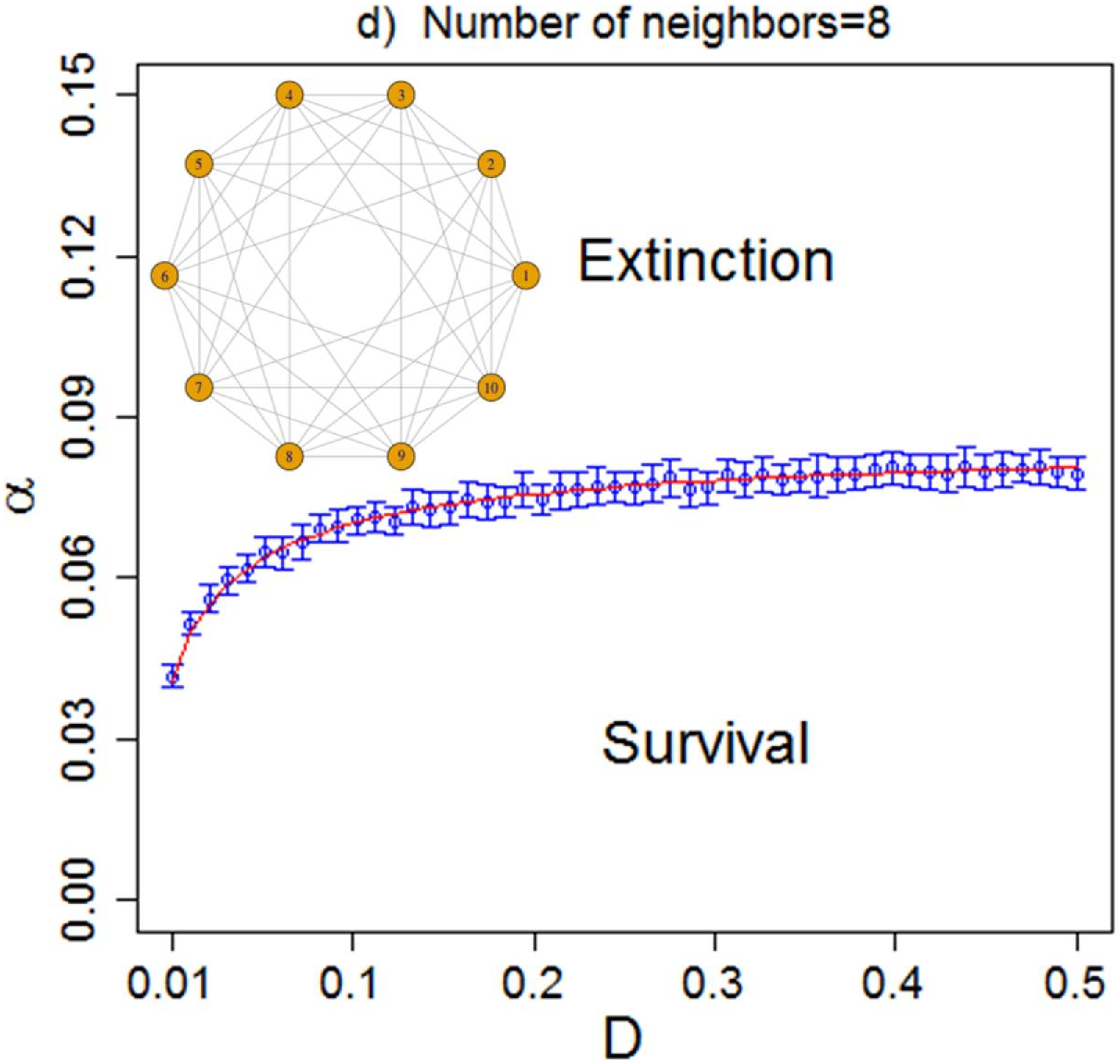}
\caption{Phase diagram $\alpha $ vs $D>0$ for networks with increasing number of neighbors $k={2,4,6,8}$  (decreasing spatial constraints). The theoretical lines (red) comes from numerical integration of Eqs \ref{Eq:Vu}-\ref{Eq:Au}.  }
\label{Fig:alpha_vs_D_k}
\end{figure}

\begin{figure}[!hbtp]
\centering
\includegraphics[width=0.5\linewidth]{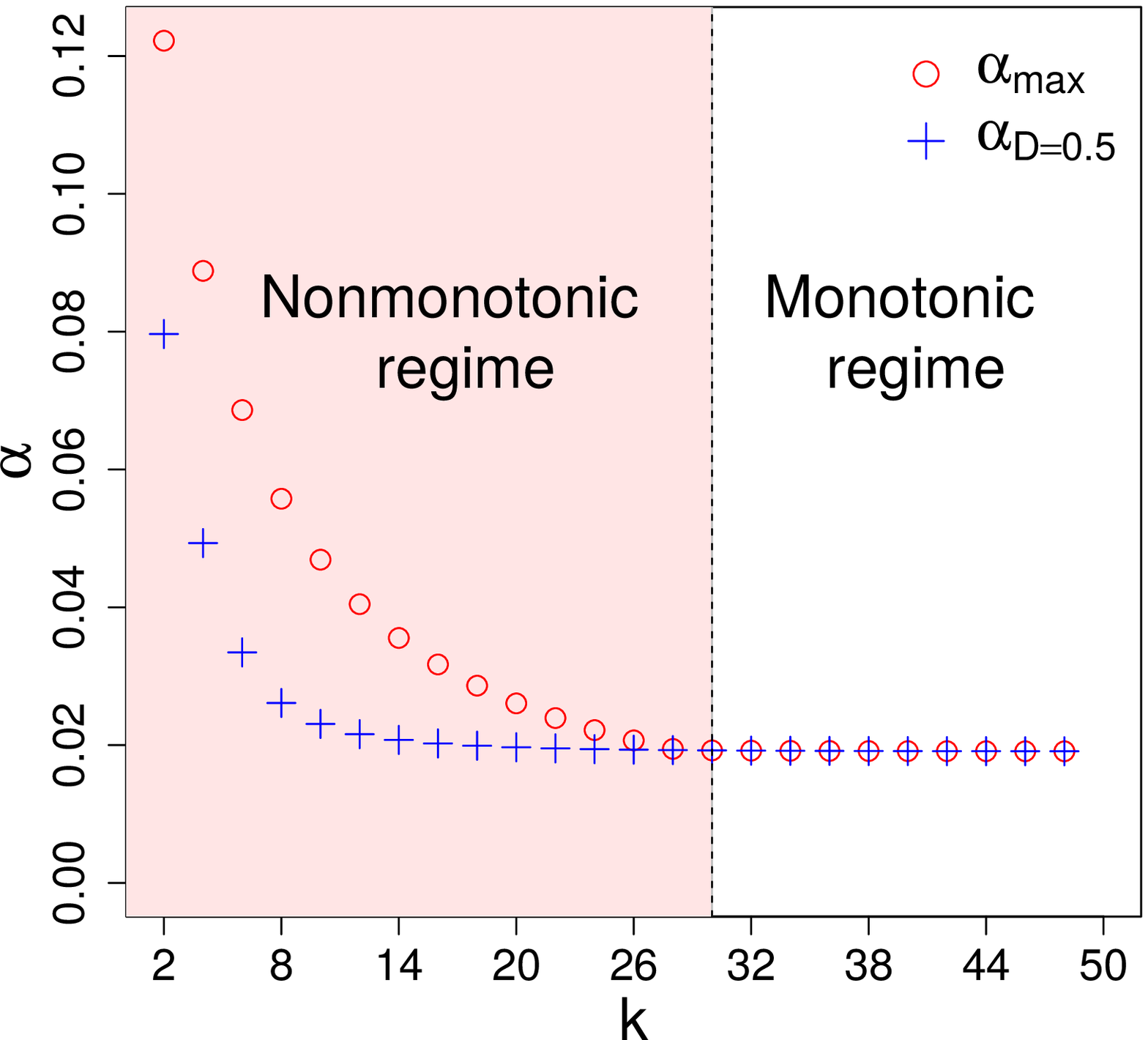}
\caption{Regime diagram of the dependence between threshold mortality $\alpha_c$ vs diffusion rate $D$ for $L=50$. The vertical line that separates the two regimes is $k_{threshold}=30$. For $k<k_{threshold}$: $\alpha_{max}>\alpha_{D=0.5}$ then $\alpha_c \times D$ displays  a nonmonotonic dependence. For $k\geq k_{threshold}$: $\alpha_{max}=\alpha_{D=0.5}$ then $\alpha_c \times D$ exhibits a monotonic dependence. }
\label{fig:phasediagramallee-net-allD-alpc-N500000-L50-fatorIo100-nSour1-1}
\end{figure}

In this section, we present our results for metapopulations of sizes $10\leq L\leq50$ and increasing $k$, but all of the results remain valid for larger networks as we checked using Monte Carlo Simulation and our coupled differential equations ($1$)-($2$), which represent the limit of very large systems. For the sake of simplicity and without losing generality for our results we fix $\lambda=1$.
%This is equivalent to rescale all the variables with the reproduction rate $\lambda$.

Let us start by looking at the time series of the total number of agents in the metapopulation for different diffusion rates as depicted in Fig.\ref{Fig:timeseries}.
The temporal evolutions for $D=\{0.03,0.09,0.019\}$ exhibit a single stable (steady) state, but the cases with $D=\{0.05,0.07,0.014\}$ display bistable solutions. This rich dynamics is the outcome of competition between the reproducibility  and mortality. 
It is worth stressing the role of randomness --- governed by our probability parameters ---  in revealing that bistability.
A clear outcome of the combination of randomness and bistability
 is the existence of ecological scenarios in which extinction can take place without apparent reason, even in the presence of abundant resources. Last, a scenario marked by two well-separated stochastically-induced steady-states is a hallmark
 of a sudden phase transition as anticipated in the previous section. Such discontinuous transition is confirmed in Fig.~\ref{Fig:Avsalpha_1} where we show the density of individuals, which is our order parameter, displays a pronounced jump for a critical mortality rate $\alpha_c$. 
To fully grasp the idea behind the survival-extinction transition in Fig.~\ref{Fig:Avsalpha_1}, consider the ecological scenarios with  $\alpha=\{0.04, 0.08, 0.12 \}$. If the environmental conditions rise the mortality from $\alpha=0.04$ to $\alpha=0.08$, 
the total density of individuals  undergoes just a slight drop 
(which may cause a false impression resilience). 
However, if the mortality increase from point $\alpha=0.08$ to $\alpha=0.12$, 
there is tremendous dynamical response in the population density namely the mass extinction. That is, the same amount of rising in mortality rate can spark either a small or drastic decline in the population.  In other words, the population can behave either in a robust or fragile manner to environmental perturbations depending on the proximity to the threshold point. 
This feature  is a remarkable fingerprint of discontinuous phase transition. It is worthwhile to mention that abrupt phase transitions are not an odd phenomenon biological dynamics~\cite{Sole_Book_2011}.

Up to now, we have not distinguish between the role of $D$ and that of $n_s$ on the threshold $\alpha_c(D)$. In order to separate out each contribution we call attention to Fig.~\ref{Fig:alpha_vs_D_ns} disentangles the role played by the interplay between the $D$ and $n_s$. To estimate the thresholds we have employed an iterative procedure quite similar to that described in section $2.1$ of  Ref.~\cite{windus2007}:
(i) first we set an initial guess for the threshold $\alpha_c{'}$, then the dynamics starts;
(ii) if a given sample enters in the extinction state we decrease $\alpha_c{'}$ by a given amount $d\alpha$;
(iii) if  a given sample has a long-term persistent population, then we increase $\alpha_c{'}$ by a given amount $d\alpha$. In the Fig.~\ref{Fig:alpha_vs_D_ns} we see that this iterative procedure provides a reasonable good estimation of the threshold that agrees very well with the theoretical threshold obtained from Eqs.~(\ref{Eq:Vu})-(\ref{Eq:Au}).
Also note that there is an optimal diffusion rate that allows the population to have comparatively high threshold mortality rates $\alpha_c$. 
The number of sources do not change the nonmonotonic dependence of $\alpha$ vs $D$, but it changes the magnitude of this dependence.

Interestingly, Fig.\ref{Fig:survival_area_1} shows there exists an optimal number of sources that promotes the largest survival area in the diagram $\alpha $ vs $D>0$, as antecipated in Fig.~\ref{Fig:alpha_vs_D_ns}; that is to say, the survival probability is maximised for an intermediate number of sources, wherefrom we understand that in populations subjected to the Allee Effect it is best to spare the population in many sources, but not too much. Similar results were found in Ref.~\cite{Zhou_ShuRong_GangWang2006} where the authors came up with an integrated model that displays an Allee-like effect at the metapopulation level, which is the outcome of imposing the  Allee effect at the local population level. That is in contrast with our  work because we use a microscopic model with no extra assumption on the birth and death rates.  

The survival-extinction phase diagram in Fig.~\ref{Fig:alpha_vs_D_k} shows that a decrease in the severity of the spatial constraints --- i.e., an increase of $k$ --- leads to a decreasing in the threshold mortality  $\alpha_c(D)$ for all $k$. That is to say, the population becomes more vulnerable to extinction when there are more open paths to emigrate. This result is supported by  Ref.~\cite{Ackleh_Allen_Carter2007} where it was found that \textit{``with fewer connections, the probability of invasion is greater''}. Furthermore, we observe the emergence of two regimes: $\alpha_c$ increases nonmonotonically with $D$ for severe spatial constraints ($k=2,4$),  but it increases monotonically with $D$ for loose spatial constraints ($k=6,8$). Although we used a simplified minimal network it already shows the importance of spatial constraints in changing the qualitative behaviour of the system. At last,  Fig.~\ref{fig:phasediagramallee-net-allD-alpc-N500000-L50-fatorIo100-nSour1-1} summarises our results for different magnitudes of spatial constraints $k$. Clearly there is a threshold for $k$, above which there is a monotononic dependence between $\alpha_c$ and $D$.

What is the  underlying mechanism behind the qualitative change presented in Fig.\ref{Fig:alpha_vs_D_k}-\ref{fig:phasediagramallee-net-allD-alpc-N500000-L50-fatorIo100-nSour1-1}?
When the geometric constraints are very severe, we have a nonmonotonic regime caused by the source-sink dynamics between the donor subpopulation and its surroundings. For small diffusion, the source cannot provide enough individuals to produce a sustainable colony in the first-neighbours that in turn acts as a drain from the donor subpopulation. For intermediate diffusion the first neighbours receive enough individuals to bear sufficient reproduction to overcome the Allee Effect. However, if the diffusion is further augmented, then the first neighbours receive as many individuals as they lose for the next-nearest neighbours, which yields an insufficient net reproduction to foster long-term survival. Alternatively, in the monotonic regime the of loose spatial constraints allows the emergence of multiple secondary sources that feed one another in a way that by boosting the diffusion one enhances the net reproduction to overcome the Allee effect.

From the empirical side, the specific  work of Smith et al  \cite{RSmith2014} supports our finding of the optimal diffusion. Therein, they engineered \textit{\it E. coli} colonies aiming at 
displaying the strong Allee effect and found that dispersal acts as a double-edged sword. In other words, intermediate dispersal rates favours  bacterial spreading whereas both low and high dispersal rates 
inhibits the spreading.

\section{Final Remarks}

We have investigated the spectrum of scenarios arising from a metapopulation dynamics under the Allee Effect using a minimal individual-based model which points at describing fundamental mechanisms thereof. Employing numerical and analytical tools we have showed that the survival-extinction boundary has a nonmonotonic behaviour for severe spatial constraints and but a monotonic behaviour for loose spatial constraints. The verification of this qualitative change in the dependence of the mortality threshold as a function of the diffusion highlights the importance of the threefold interplay between the Allee Effect, diffusion and geometric constraints for the persistence of populations.
Besides the experimental work of Ref.~\cite{RSmith2014}, there are   previous theoretical models pointing to our conclusions over the likely existence of an intermediate mobility rate that optimises the survival probability. Explicitly, Ref.~\cite{Yang_ZXWu_Holme_Nonaka2017} found a \textit{``a nonmonotonic dependence of the critical Allee thresholds on the migration rate.''} by imposing the Allee Effect at the microscopic scale considering a nonlinear per capita birth
rate $rn_i/C + rn_ic/C^2$ per capita death rate 
$rn_i^2/C + rc/C$.\footnote{$n_i$ stands for the number of individuals on habitat patch $i$, $C$ is the carrying capacity, $c$ is an  Allee threshold} In addition, we can also refer to Ref.~\cite{South_Kenward2001} in which it was used an individual two-gender population on a hexagonal grid where the juveniles disperse away from their natal territory with dispersal distances distributed as a negative exponential. In that case, the population growth was highest for an optimal distance of the dispersal. Yet, both works did not observe the fact that the magnitude of the spatial constraints can change qualitatively the survival-extinction boundary from a nonmonotonic to a mononotonic dependence. 
Our finding prompts an inquiry into the actual role of network topology in the macroscopic outcome of ecological dynamics; something we intend to explore in future work.

In a broader view, there are other biological systems that exhibit nonmonotonic effects of diffusion such as epidemic spreading~\cite{Silva2018}, birth-death-competition dynamics with migration~\cite{Lampert_Hastings2013}, evolutionary dynamics with the Allee effect and sex-biased dispersal~\cite{Shaw_Kokko2015}, logistic growth dynamics in metapopulations with heterogeneous carrying capacities~\cite{Khasin2012}, metapopulation genetics dynamics with balancing selection~\cite{Lombardo2014},
two-type (mutants, strains, or species) population dynamics under the Allee effect~\cite{Korolev2015}, and range expansion of a genetically diverse population where  individuals may invest its limited resources partly in motility and partly in reproduction~\cite{Reiter2014}.
As we adopted a minimal ecological model, it is possible to bring forth different extensions of the present work in order to fit for the traits of the problems we have just mentioned.
For instance, instead of using a memoryless random walk, we can use a more realistic mobility dynamics: random walks that intermittently revisits previously visited places \cite{Boyer2014}. 

%Another promising avenue of research is ...

\vspace*{1.5cm}

%\subsection*{Acknowledgments}

$\bullet$ Acknowledgments

The authors acknowledge financial support from the Brazilian funding agencies CAPES (MAP) as well as CNPq and FAPERJ (SMDQ).

$\bullet$ Competing interests

The authors have declared that no competing interests exist.

%\subsection*{Authors' contributions}

$\bullet$ Authors' contributions

Conceptualization: MAP and SMDQ; Formal analysis: MAP; Funding acquisition: SMDQ; Investigation: MAP and SMDQ; Methodology: MAP and SMDQ; Software: MAP; Supervision: SMDQ; Validation: MAP and SMDQ; Writing - original draft: MAP; Writing - review \& editing: SMDQ.

%\clearpage

\bibliographystyle{plain}

\end{document}